\documentclass[nato,noid,numreferences]{crckbked}
\usepackage{graphicx}
\usepackage{epsf}
\makeatletter
\def\fnum@figure{\relax}
\makeatother

\begin{document}
\begin{opening}
\title{DYNAMICAL MASS GENERATION IN QUANTUM FIELD THEORY : SOME METHODS WITH APPLICATION TO THE GROSS-NEVEU MODEL AND YANG-MILLS THEORY\protect}
\author{david dudal \footnote{david.dudal@rug.ac.be} \footnote{Research Assistant of the Fund For Scientific
Research-Flanders (Belgium)}}
\author{karel van acoleyen\\}
\author{henri verschelde}
\institute{Ghent University\\
           Department of Mathematical
Physics and Astronomy\\Krijgslaan 281-S9 \\
B-9000 GENT, BELGIUM}

\begin{abstract}
We introduce some techniques to investigate dynamical mass
generation. The Gross-Neveu model \cite{Gross} (GN) is used as a
toy model, because the GN mass gap is exactly known
\cite{Forgacs}, making it possible to check reliability of the
various methods. Very accurate results are obtained. Also
application to SU($N$) Yang-Mills (YM) is discussed.
\end{abstract}

\end{opening}

\section{Introduction}
Recently, there was growing evidence that the YM-vacuum favours a
condensate of mass dimension 2 \cite{Gubarev1,Gubarev2}. A
reasonable candidate is
\begin{equation}\label{1}
    \Delta=\frac{1}{2}(VT)^{-1}\left\langle \min_{U}\int
    d^{4}x\left(A_{\mu}^{U}\right)^{2}\right\rangle
\end{equation}
where $U$ labels an arbitrary gauge transformation. (\ref{1}) is a
gauge invariant operator and so has some physical meaning. It
reduces to the local composite operator (LCO)
$\frac{1}{2}\left\langle A_{\mu}^{2}\right\rangle$ in the Landau
gauge. The Landau gauge is in a way the most natural one to
perform calculations. With another gauge fixing, (\ref{1}) is no
longer local. More precisely, thinking of Abelian projection and
imposing the Maximally Abelian Gauge (MAG), we should consider the
gauge invariant operator $\Delta=\frac{1}{2}(VT)^{-1}\left\langle
\min_{U}\int d^{4}x \left(A_{\mu}^{a}A^{\mu
a}\right)^{U}\right\rangle$ where the index $a$ runs only over the
off-diagonal gluons. In MAG, this simplifies to the LCO
$\frac{1}{2}\left\langle A_{\mu}^{a}A^{\mu a}\right\rangle$.
\\\\We mentioned MAG, because interesting attempts have been
undertaken by Kondo e.a. \cite{Kondo1} and recently Freire
\cite{Freire} to construct low energy effective theories starting
from the full YM Lagrangian. Their efforts rest mainly on the
principle of Abelian dominance \cite{Suzuki,Amemiya}, which can be
understood by means of massive off-diagonal gluons and the
Appelquist-Carazzone decoupling theorem \cite{Apple}. MAG requires
a 4-ghost interaction to preserve renormalizibility \cite{Min},
and this non-trivial ghost interaction was used in
\cite{Kondo2,Schaden} to produce an effective charged gluon mass.
Our analysis learned that, in contrast to their results, the ghost
condensate alone resulted in a tachyonic mass. We guess that a
combination\footnote{Some more formal results on this topic were
obtained in \cite{Kondo3}.} of the off-diagonal ghost and gluon
condensate might provide us with a real mass. \\\\For the sake of
simplicity, we present the different approaches using the
GN-model. Available results on YM in the Landau gauge will be
quoted too.
\section{Renormalizable effective potential for LCO}
\subsection{Gross-Neveu}
We start from the manifestly U($N$)-invariant GN Lagrangian in
$2-\varepsilon$ dimensional Euclidean spacetime with a source $J$
coupled to the LCO $\overline{\psi}\psi$
\begin{equation}\label{3}
    {\cal L}=\overline{\psi}\left(\partial \hspace{-0.2cm}/ +J\right)\psi-\frac{1}{2}g^{2}\mu^{\varepsilon}\left(\overline{\psi}\psi\right)^{2}+{\cal L}_{counter}
\end{equation}
This is an asymptotically free theory with a chiral $\gamma_{5}$
symmetry if $J=0$, preventing a perturbative non-zero value for
$\left\langle\overline{\psi}\psi\right\rangle$ and the fermion
mass. When $J\neq0$, new logarithmic infinities $\propto J$
(multiplicative mass renormalization) and $\propto J^{2}$ (vacuum
energy divergences) appear. In order to remove the latter, a new
coupling $\zeta$ must be incorporated into (\ref{3}), and we get
\begin{equation}\label{4}
    {\cal L}=\overline{\psi}\left(\partial \hspace{-0.2cm}/ +J\right)\psi-\frac{1}{2}g^{2}\mu^{\varepsilon}\left(\overline{\psi}\psi\right)^{2}-\frac{1}{2}\mu^{-\varepsilon}\zeta J^{2}+{\cal L}_{counter}
\end{equation}
where ${\cal L}_{counter}$ contains all counterterm information.
\begin{equation}\label{5}
    {\cal L}_{counter}=\delta Z\overline{\psi}(\partial \hspace{-0.2cm}/)\psi+\delta Z_{2}J\overline{\psi}\psi-\frac{1}{2}\delta Z_{g}g^{2}\mu^{\varepsilon}\left(\overline{\psi}\psi\right)^{2}-\frac{1}{2}\mu^{-\varepsilon}\delta\zeta J^{2}
\end{equation}
Defining the bare quantities as
\begin{eqnarray}
    \psi_{o} &=& \sqrt{Z}\psi  \\
    J_{o} &=& \frac{Z_{2}}{Z}J  \\
    g_{o}^{2} &=& \frac{Z_{g}}{Z^{2}}g^{2}\\
    \zeta_{o}J_{o}^{2} &=& \mu^{-\varepsilon}(\zeta + \delta\zeta)J^{2}
\end{eqnarray}
the finite, generating energy functional $E(J)$ satisfies a
homogeneous renormalization group equation (RGE)
\begin{equation}\label{6}
    \left(\mu\frac{\partial}{\partial\mu}+\beta\left(g^{2}\right)\frac{\partial}{\partial g^{2}}-\gamma_{2}\left(g^{2}\right)\int
    d^{2}xJ\frac{\delta}{\delta J}+\eta\left(g^{2},\zeta\right)\frac{\partial}{\partial\zeta}\right)E = 0
\end{equation}
with
\begin{eqnarray}\label{7}
    \beta\left(g^{2}\right)&=&\left.\mu\frac{\partial}{\partial\mu}g^{2}\right|_{g_{o},\varepsilon}\\
    \gamma_{2}\left(g^{2}\right)&=&\left.\mu\frac{\partial}{\partial\mu}\ln\frac{Z_{2}}{Z}\right|_{g_{o},\varepsilon}\\
    \eta\left(g^{2},\zeta\right)&=&\left.\mu\frac{\partial}{\partial\mu}\zeta\right|_{g_{o},\varepsilon,J_{o},\zeta_{o}}
\end{eqnarray}
The above reasoning to treat a LCO seems to have 2 problems :
First, $\zeta$ is at this stage still arbitrary, so we have a
problem of uniqueness. Putting $\zeta=0$ is a bad choice, it leads
to a non-homogeneous RGE for $E$ and a non-linear RGE for the
effective action $\Gamma$. Secondly, the $J^{2}$ term spoils a
simple energy interpretation. These are reflections of the
criticism uttered by Banks and Raby on the use of LCO
\cite{Banks}. Both problems can be solved by choosing $\zeta$ such
a function of $g^{2}$, so that if $g^{2}$ runs according to
$\beta$, $\zeta$ will run properly according to $\eta$
\cite{Verschelde1}. Indeed, because of (\ref{6}) and (\ref{7}), we
have
\begin{equation}\label{8}
    \mu\frac{\partial}{\partial\mu}\zeta = \eta =
    2\gamma_{2}\zeta+\delta
\end{equation}
where
\begin{equation}\label{9}
   \delta =
    \varepsilon\delta\zeta-\mu\frac{\partial}{\partial\mu}\delta\zeta+2\gamma_{2}\delta\zeta
\end{equation}A solution of (\ref{8}) is $\zeta=\zeta(g^{2})$ where
$\zeta(g^{2})$ is a particular solution of
\begin{equation}\label{11}
    \beta(g^{2})\frac{d}{dg^{2}}\zeta(g^{2})=2\gamma_{2}(g^{2})\zeta(g^{2})+\delta(g^{2})
\end{equation}The integration constant has been put to zero, in order to avoid
an independent coupling constant and to have multiplicatively
renormalizable vacuum divergences $\left(
\zeta+\delta\zeta=Z_{\zeta}\zeta\right)$. We will solve (\ref{11})
by a Laurent expansion $
\zeta=\frac{z_{-1}}{g^2}+z_{0}+z_{1}g^{2}+\ldots$ Notice that
$n$-loop results require $(n+1)$-loop knowledge of
$\beta$,$\gamma_{2}$ and $\delta$. The generating functional $E$
now fulfills
\begin{equation}\label{13}
    \left(\mu\frac{\partial}{\partial\mu}+\beta\left(g^{2}\right)\frac{\partial}{\partial
g^{2}}-\gamma_{2}\left(g^{2}\right)\int
    d^{2}xJ\frac{\delta}{\delta J}\right)E = 0
\end{equation}We conclude that the LCO $\Delta=Z_{2}\overline{\psi}\psi-Z_{\zeta}\zeta
J$ has a finite and multiplicatively renormalizable VEV
$\langle\Delta\rangle = \frac{\delta W}{\delta J}$. The effective
action $\Gamma(\Delta)$, obeys the following RGE
\begin{equation}\label{16}
\left(\mu\frac{\partial}{\partial\mu}+\beta\left(g^{2}\right)\frac{\partial}{\partial
g^{2}}+\gamma_{2}\left(g^{2}\right)\int
    d^{2}x\Delta\frac{\delta}{\delta \Delta}\right)\Gamma(\Delta) = 0
\end{equation}
Introducing unity via the Hubbard-Stratonovich transformation
\begin{equation}\label{17}
    1=\int[d\sigma]\exp-\frac{1}{2Z_{\zeta}\zeta}\int
    d^{2-\varepsilon}x\left[\frac{\sigma}{g}+\mu^{\frac{\varepsilon}{2}}Z_{2}\overline{\psi}\psi-\mu^{\frac{-\varepsilon}{2}}Z_{\zeta}\zeta
    J \right]^2
\end{equation}
we finally arrive at
\begin{equation}\label{18}
    \exp(-E(J))=\int[d\overline{\psi}d\psi d\sigma]\exp-\int
    d^{2-\varepsilon}x\left[{\cal L}(\sigma,\overline{\psi},\psi)-\mu^{\frac{-\varepsilon}{2}}\frac{\sigma}{g}J\right]
\end{equation}
where
\begin{equation}\label{19}
    {\cal
    L}=Z\overline{\psi}\partial\hspace{-0.2cm}/\psi-\frac{1}{2}\mu^{\varepsilon}g^{2}\left(\overline{\psi}\psi\right)^{2}\left[Z_{g}-\frac{Z_{2}^{2}}{g^{2}Z_{\zeta}\zeta}\right]+\frac{\sigma^{2}}{2g^{2}Z_{\zeta}\zeta}+\mu^{\frac{\varepsilon}{2}}g\sigma\overline{\psi}\psi\frac{Z_{2}}{g^{2}Z_{\zeta}\zeta}
\end{equation}
$J$ is now a real source, in the sense that it appears linearly
for $\sigma$ so that we have a straightforward energy
interpretation and $\langle\sigma\rangle =\langle-g\Delta\rangle$.
Eq.(\ref{19}) is a new effective, renormalized Lagrangian for GN,
equivalent to the original (\ref{3}) but encapsulating
non-perturbative information. Perturbing around $\sigma=0$, we
recover the original perturbation series with its infrared
renormalon problems. If we calculate the effective potential for
$\sigma\neq0$, we could perturb around a non-perturbative vacuum
free of renormalons. \\\\We won't rederive $V(\sigma)$, the
results can be found in the original papers \cite{Verschelde1}.
Out of (\ref{19}), we immediately see that
$\langle\sigma\rangle\neq0$ gives birth to a fermion mass. After
improving the renormalization prescriptions, very accurate results
for the effective fermion mass were obtained (see TABLE I). For
comparison, we also displayed the $N\rightarrow\infty$ and $1/N$
results.
\begin{center}\label{tabel1}
\begin{table}[htb]
\caption{Deviation in terms of percentage for the mass gap with
LCO method}
\begin{tabular}{cccc}
\hline
$N$ & 2-loop mass gap & $N\rightarrow\infty$ mass gap &$1/N$ mass gap\\
\hline
2& 41.67\% & -46.3\%& -21.9\%\\
3& 7.13\% & -32.5\%& -12.2\%\\
4& 2.84\% & -24.2\%& -7.0\%\\
5& 1.53\% & -19.1\%& -4.5\%\\
6& 0.97\% & -15.8\%& -3.1\%\\
7& 0.68\% & -13.5\%& -2.3\%\\
8& 0.51\% & -11.7\%& -1.8\%\\
9& 0.40\% & -10.4\%& -1.4\%\\
10& 0.33\% & -9.3\%& -1.1\%\\
\hline
\end{tabular}
\end{table}
\end{center}
\subsection{Yang-Mills with Landau gauge fixing}
The same approach has been employed to investigate the condensate
suggested in (\ref{1}). There are some subtleties since YM is a
gauge theory \cite{Verschelde2}. The corresponding effective
YM-Lagrangian, was found to be
\begin{eqnarray}\label{20}
    {\cal
    L}(\sigma,A_{\mu})&=&\frac{1}{4}F_{\mu\nu}^{a}F_{\mu\nu}^{a}+{\cal
    L}_{gauge+F.P.}+{\cal
    L}_{counter}+\frac{\sigma^{2}}{2g^{2}Z_{\zeta}\zeta}\\ \nonumber &+&\frac{1}{2}\mu^{\frac{\varepsilon}{2}}g\sigma
    A_{\mu}^{a}A_{\mu}^{a}\frac{Z_{2}}{g^{2}Z_{\zeta}\zeta}+\frac{1}{8}\mu^{\varepsilon}\frac{Z_{2}^{2}}{Z_{\zeta}\zeta}\left(A_{\mu}^{a}A_{\mu}^{a}\right)^{2}
\end{eqnarray}
$V(\sigma)$ was computed up to 2-loop order using the
$\overline{MS}$ scheme. RG-improved perturbation theory showed
that all gluons are massive :
\begin{equation}\label{21}
    m_{gluon}\approx465\textrm{ MeV} \textrm{ with }    \frac{g^{2}N}{16\pi^{2}}\approx0.14466
\end{equation}
Note that the relevant expansion parameter is relatively small, so
perturbation theory can be qualitatively trusted.

\section{Dynamical mass generation by source inversion}
The second tool we discuss, was worked out recently by Van
Acoleyen et al. \cite{Karel1}. \\\\When the GN model is probed
with a source $J$, just as in (\ref{3}), one can calculate the
effective mass $m(J)$ as a function of $J$. Due to the asymptotic
freedom, this expansion is only valid for large $J$. To recover
the original GN model, we must take the limit $J\rightarrow0$.
Doing so, the perturbation
series for $m(J)$ blows up and no relevant information can be extracted.\\
However, it is possible to invert the relation $m(J)$ to $J(m)$.
If a sufficiently large solution $m_{\star}$ of $J(m)=0$ exists,
we can consider the limit of vanishing source, while the
perturbative expansion remains valid.\\\\
The solution $m_{\star}\neq0$ will be renormalization scheme and
scale dependent, due to the arbitrary renormalization
prescriptions. To remove this freedom, we proceed in the following
way. $J$ runs according to its renormalization group equation as
\begin{equation}\label{60}
    \mu\frac{\partial J}{\partial\mu}=-\gamma_{2}\left(g^{2}\right)J
\end{equation}
$J$ is a scheme and scale dependent quantity, with the result that
the equation for the mass gap is also scheme and scale dependent.
But it is easily checked that $\tilde{J}$, defined by
\begin{equation}\label{61}
    \tilde{J}=f\left(g^{2}\right)J
\end{equation}
where $f\left(g^{2}\right)$ is a solution of
\begin{equation}\label{62}
    \mu\frac{\partial f}{\partial\mu}=\gamma_{2}f
\end{equation}
is scheme and scale independent (SSI).\\ When we transform $J$ to
$\tilde{J}$, the gap equation becomes $\tilde{J}(m)=0$, since
$\tilde{J}\propto J$.  Because $\tilde{J}$ is SSI, the gap
equation is SSI, so $m_{\star}$ will be SSI.\\\\
Since we can calculate the perturbative series for $m(J)$ an
$\tilde{J}$ only up to a certain order, there will always be a
remnant of scheme and scale dependence. By exchanging the
expansion parameter $g^{2}(\mu)$ for
$\frac{1}{\beta_{0}\ln\frac{\mu^{2}}{\Lambda^{2}}}$ ($\Lambda$ is
the scale parameter of the renormalization scheme), it is possible
to rewrite $\tilde{J}$ as $\tilde{J}=m{\cal
J}\left(\frac{m}{\Lambda_{\overline{MS}}}\right)$, where ${\cal
J}$ is a series in
$\frac{1}{\beta_{0}\ln\frac{m^{2}}{\Lambda_{\overline{MS}}^{2}}+d}$
with all scheme and scale dependence settled in the parameter $d$.
\\\\When we would include all orders in the calculation, $d$ would
drop out of the result. At finite order precision, $d$ will be
present in the final result for $m_{\star}$. We can fix $d$ by
using the \emph{principle of minimal sensitivity} (PMS) \cite
{Stevenson} by demanding that the mass gap $m_{\star}$ has minimal
dependence on $d$ $\left(\Leftrightarrow\frac{\partial
m_{\star}}{\partial d}=0\right)$. At 2 loop order, PMS did give an
optimal $d$ and the corresponding mass was close to the exact mass
(see TABLE II).
\begin{center}\label{tabel2}
\begin{table}[htb]
\caption{Deviation in terms of percentage for the mass gap with
source inversion method}
\begin{tabular}{cccc}
\hline
$N$ & 2-loop mass gap & $N\rightarrow\infty$ mass gap &$1/N$ mass gap\\
\hline
2& $\pm20$\% & -46.3\%& -21.9\%\\
3& 0.9\% & -32.5\%& -12.2\%\\
4& -1.0\% & -24.2\%& -7.0\%\\
5& -1.5\% & -19.1\%& -4.5\%\\
6& -1.6\% & -15.8\%& -3.1\%\\
7& -1.6\% & -13.5\%& -2.3\%\\
8& -1.5\% & -11.7\%& -1.8\%\\
9& -1.4\% & -10.4\%& -1.4\%\\
10& -1.3\% & -9.3\%& -1.1\%\\
\hline
\end{tabular}
\end{table}
\end{center}
The source inversion method was also tested on the chiral
Gross-Neveu model, again with good results \cite{Karel2}.
\section{Summary}
We have dealt with 2 different approaches concerning dynamical
mass generation. Each method seems to give (very good) results in
case of GN. The almost exact GN results doesn't mean we can
generalize immediately to the YM case (in the Landau gauge). Other
sources of non-perturbative effects besides infrared renormalons,
such as instantons, will contribute to the dynamical mass.\\\\In
the future, also MAG YM deserves our attention, first trying to
clear the widely accepted Abelian dominance in a somewhat
analytical way. Secondly, there might exist a connection between
those condensate formations and the Curci-Ferrari Lagrangian
\cite{CF,Gracey}. \\\\ As a general conclusion, we state it is
possible to get non-perturbative information on e.g. a dynamical
gluon mass in a sector of the YM-vacuum accessible to perturbation
theory.

\end{document}